\documentclass[10pt]{article}

\usepackage{amsmath}
\usepackage{amssymb}

\usepackage{graphicx}

\usepackage{cite}
\usepackage{color} 
\usepackage{authblk}

\usepackage[labelfont=bf,labelsep=period,justification=raggedright]{caption}

\makeatletter
\renewcommand{\@biblabel}[1]{\quad#1.}
\makeatother

\date{}

\usepackage{lscape}

\begin{document}

\begin{flushleft}
{\Large
\textbf{The Accounting Network: how financial institutions react to systemic crisis}
}
\\
Michelangelo Puliga$^{1,2.\ast}$, 
Andrea Flori$^{1}$, 
Giuseppe Pappalardo$^{1}$
Alessandro Chessa$^{1,2}$
Fabio Pammolli$^{1}$
\\
\bf{1} IMT, School for Advanced Studies, Lucca, Italy
\\
\bf{2} Linkalab, Complex Systems Computational Laboratory, 09129 Cagliari, Italy
\\
\bf{*}{\small{Correspondence and requests for materials should be addressed to M.P. (\emph{michelangelo.puliga@imtlucca.it})}}

\end{flushleft}

\vspace{0.5cm}

\vspace{0.5cm}


\section*{Abstract}
The role of Network Theory in the study of the financial crisis has been widely spotted in the latest years. It has been shown how the network topology and the dynamics running on top of it can trigger the outbreak of large systemic crisis. Following this methodological perspective we introduce here the Accounting Network, i.e. the network we can extract through vector similarities techniques from companies' financial statements. We build the Accounting Network on a large database of worldwide banks in the period 2001-2013, covering the onset of the global financial crisis of mid-2007. After a careful data cleaning, we apply a quality check in the construction of the network, introducing a parameter (the Quality Ratio) capable of trading off the size of the sample (coverage) and the representativeness of the financial statements (accuracy). We compute several basic network statistics and check, with the Louvain community detection algorithm, for emerging communities of banks. Remarkably enough sensible regional aggregations show up with the Japanese and the US clusters dominating the community structure, although the presence of a geographically mixed community points to a gradual convergence of banks into similar supranational practices. Finally, a Principal Component Analysis procedure reveals the main economic components that influence communities' heterogeneity. Even using the most basic vector similarity hypotheses on the composition of the financial statements, the signature of the financial crisis clearly arises across the years around 2008. We finally discuss how the Accounting Networks can be improved to reflect the best practices in the financial statement analysis.  



\section*{Introduction}
Network Theory has been used to establish how contagion, through a variety of channels (mutual exposures, social networks of board members, moral hazard from permissive regulations, financial instruments like swaps and derivatives, etc.), triggered the outbreak of the 2007-08 crisis. Scholars suggest that financial systems may affect positively economic development and its stability (\cite{Beck09, Beck11, Lev05}), although they may represent a source of distress which leads to bank failures and currency crises, or greater contraction for those sectors that depend more on external finance during banking crisis (\cite{arr, RR}). As a response to the recent financial turmoil, the banking sector has been affected by a substantial reorganization (\cite{BISANNUAL14}). For instance, as highlighted by the European Central Bank for the Euro area \textit{the main} \textit{findings reflect the efforts by banks to rationalize banking businesses, pressure to cut costs, and the deleveraging process that the banking sector has been undergoing since the start of the financial crisis in 2008} (\cite{ECB}). This implies that market pressure and regulatory amendments induce banks to reduce their levels of debt, through cost containment and stricter capital requirements. In addition, a gradual improvement in bank capital positions aims to enhance the capacity of the system to absorb shocks arising from financial and economic distresses. This limits the risk of spillover effects from the financial sector to the real economy and put the financial system in a better condition to reap the benefits of economic recovery. In particular, as the financial boom turned to a bust, banks' stability deteriorated abruptly and the economy entered a \textit{balance sheet recession}, which depressed spending levels through a reduction in consumption by households and investments by firms. Therefore, although at an uneven pace across regulations, the need to strengthen fundamentals has influenced the banking sector, and differences in banks' portfolio allocations, financial performances, and capitalizations might be interpreted as the combined results of policy decisions and sectoral responses to changes in the regulatory framework (see e.g. \cite{ALLEN}, \cite{DIARANJ}).\\
\noindent This paper relates to the literature on banking development and performance evaluation during the recent crisis (see e.g. \cite{ASHIN}, \cite{BERBOW},\cite{BRUNN}). We consider a large data set of worldwide banks retrieved from \textit{Bloomberg}, focusing on financial statements spanning from 2001 to 2013. We introduce a network based on similarities between banks' financial statement compositions (hereinafter \textit{Accounting Network}). Due to data limitations, the reference sample is restricted to banks for which a continuum and stable set of variables is available for the entire period. The introduction of a methodology (\textit{Quality Ratios}) to measure banks' data coverage aims to prevent that missing values for some variables or lack of annual financial statements for some banks affect the overall picture. We then exploit the maximum amount of available information from financial statements without further reducing the set of variables through an arbitrary selection of the financial statements fields. This choice aims to avoid any selection bias. Moreover, total assets (as a proxy for size) for each bank is applied to normalize banks' financial statements measures to prevent the emergence of ``size effects'' as the sizes of institutions are spanning for various orders of magnitude.\\
The introduction of Accounting Networks establish a bridge between the external perspective arising from market data and the internal one based on banking activities indicators. We study how Accounting Networks can be exploited to provide a description of the banking system during the crisis. This part sheds light on whether banks under different regulatory frameworks and diversification degrees have reacted to the crisis by strengthening their business peculiarities or by converging towards similar practices (\cite{BelStu},\cite{DeHui}). We rely on the assumption that market data alone, although highly representative of investors' perception of the banking sector, might be dis-informative during periods of distressed market conditions. This, in turn, stimulates a broader exploitation of the information on banking activities, thus pointing to a more comprehensive investigation which takes into account also the internal perspective arising from financial statements data. In addition, the use of accounting data allows a partition of business activities where banks are involved in, providing therefore an approximation of the state of the system related to several potential channels through which the financial distress might propagate. This is appealing also for regulators, since authorities are interested in a wide set of economic indicators in order to prevent the systemic relevance of financial institutions and they introduce specific requirements and constraints which affect directly financial statements measures. For these reasons, we believe that enriching the debate on financial stability by means of the Accounting Networks might give new clues about the resilience of the banking system. \\ 
Another important result is the possibility of getting a neutral partition of banks in "network communities" (i.e. clusters) that results from the analysis of the network through community detection algorithms like the \textit{Louvain} modularity maximization. The results indicates that regional communities evolve in time and the crisis has a clear role in weakening geographically determined structures. Furthermore, we focus on proxies for leverage, size and performance in order to understand if these variables have played a key role among the set of economic measures usually applied to classify banks (see e.g. \cite{BLU},\cite{HUI}). Hence, we aim to answer the question whether the collapse of financial markets has weakened these relationships, limiting therefore the power of traditional indicators to identify clusters of homogeneous banks. Correlation diagrams applied to show how network variables are related to economic measures suggest a turning point in correspondence of the outbreak of the crisis, which influenced the role of proxies for leverage, size or performance to group similar banks. This preliminary results motivated the last section, where by means of Principal Component Analysis we investigate which economic features are more likely to characterise the heterogeneity of the communities before, during and after the collapse of 2007-08.\\
The remaining part of the work discusses open issues and future lines of research, such as open questions on how to improve the building of the Accounting Networks. In particular, the effectiveness of this approach can be enhanced by means of a careful variable selection based on the best financial practices applied in the evaluation of the financial statements structures. In addition, a more accurate normalization of the variables and caring about national regulations may increase the usefulness of the methodology. Furthermore, matrix filtering techniques and missing data reconstruction for financial statements information can enhance the extraction of meaningful clusters. Then, more advanced and focused tools could be conceived to analyse banks evolution towards similar business configurations or, alternatively, their divergent patterns as a response to changing market conditions.

\section*{Methods}
\subsection*{Dataset preparation}
\begin{figure}[h]
\centerline{
\includegraphics[scale=0.15]{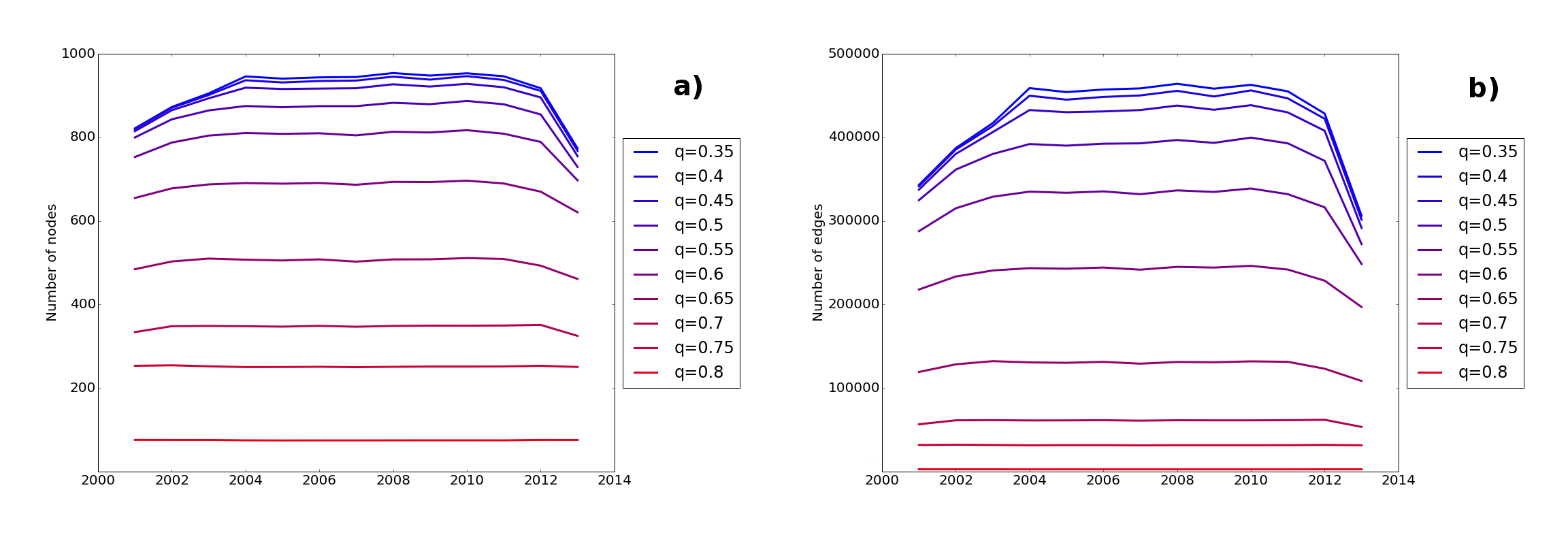}
}
\caption{This picture shows the number of nodes and edges along the sample period for different QR values. It is clear to see how for small values of the Quality Ratio parameter the curves belong to a stricter range.}
\label{figQR}
\end{figure}

The dataset we analysed covers the set of banks provided by \textit{Bloomberg} which were active (i.e. with traded instruments) at the end of the first quarter of 2014. Although quarterly information is available, we prefer to focus on annual balance sheets and income statements for accounting standard reasons, as different countries can have different obligations in terms of the provision of quarterly financial statements and this can lead to a mismatch and a poor variables coverage. Data are collected during the reference period from 2001 to end of 2013.\\ 
As regards financial statements data, we select a large set of variables among those available in \textit{Bloomberg} and related to the current regulatory framework (\cite{BASEL}). We rely on the existing literature for the selection process, although providing a neutral approach. We focus the analysis on proxies for banking business models (see e.g. \cite{ALTU},\cite{CAL}). In particular, balance sheet data provide a year-by-year picture of stock variables in terms of assets and liabilities for different instruments and maturities, while income statement data describe annual economic performances by partitioning profits and losses according to banking activities ranging for instance from interests to fees. Since national regulations allow firms to fix a different end of fiscal year, we extend the ``end of year'' definition and the relative financial statements according to a window in the range between three months before and after the end of the solar year. Solving overlapping issues in variables definitions, as well as the base currency choice, constitute the first step in the data pre-processing procedure. Firstly we discard total and sub-total measures (as they are redundant measures), and secondly we choose US dollars as currency base, thus facilitating banks comparisons.\\
Working with financial statements data often leads to limitations in data coverage and completeness. Therefore, the starting point of our analysis is represented by the selection of a stable set of banks in terms of data availability during the sample period. In particular, banks might change the composition of their financial statements or they might be excluded by the \textit{Bloomberg} provider due to several reasons, such as for instance a new regulation or a change in the bank's economic activities. This, in turn, might cause \textit{missing values} for some variables or lack of financial statements for several banks in certain years. In order to limit the impact of these issues on our findings, we define a methodology to measure the coverage of available variables for each bank in the reference period. We refer to the \textit{Quality Ratios (QRs)} as the proportion of available and usable variables $V_{OK}$ over the maximum of all possible ones $V_{ALL}$ in the sample period: $QR=V_{OK}/V_{ALL}$. The tuning of this indicator, combined with two more filters on the frequency of financial reporting, provides a stable set of banks identified by their QR. The two additional criteria are: a minimum number of financial statements of ten out of thirteen possible fiscal years and a maximum gap period between two consecutive annual reports equal to seven hundred days. Once selected those banks that report almost continuously their financial statements, we study them according to their respective QR.\\ Actually, individual QRs, as empirically computed on the entire perimeter, lie in the range between 0.3 (low accuracy/coverage) and 0.8 (high accuracy/coverage). Interestingly, many measures computed on the sets of banks obtained by fixing the QR do not seem to be significantly affected by its choice (except, as expected, for high QRs, where the size of the sample reduces significantly). With greater values of the QR parameter we have less available banks to be considered, since only few of them have a large set of variables present in many of their financial statements. As the estimates are stable in a reasonable QR range, in this work we decide to use the set arising from the case of QR = 0.5 that, even if arbitrary, still represents a good compromise between the accuracy of the estimate and the size of the sample (see Figure \ref{figQR}).\\

\subsection*{Accounting networks}
For every year a vector of financial statement variables is assigned to each bank and used to compute the cosine similarities between pairs of banks/nodes. Here the intuition is that the most similar banks (as from their financial statements) must stay closer in the network and form a cluster. Then, the measure ``cosine similarity'' is transformed into a metrics (as triangular inequality must hold, the square root is used). The definition is the following: we compute the cosine of the angle between each pair of vectors with the dot product and then we apply the simple transformation $w_{i,j} = 1 - \sqrt{1 - C_{i,j}^2}$, where $w_{i,j} \in [0, 1]$ and $C_{i,j}$ is the cosine similarity between \textit{i} and \textit{j}. In network terms $w_{i,j}$ is the weight. This transformation (see \cite{DongenEnright} for an introduction to similarity measures and relative metrics) is used to move from the cosine similarities defined in the space [-1,1] to weights in the interval [0,1]. With this transformation the more two nodes are similar (or anti-similar) the larger is the weight, while a weight of 0 is assigned to a pair of nodes having totally dissimilar financial statements (actually, in our networks cosine similarities range mainly between 0 and +1).\\
\noindent In addition, before the computation of the metrics, we need to take care of the size distribution of banks, as it spans over several orders of magnitude. To avoid a bias toward large institutions, for each bank we divide all variables in its vector by the respective total assets in such a way that the attributes of the vector refer to economic and financial \textit{ratios}. This operation ensures that clusters will be formed by banks with similar business activities regardless their sizes.\\ 
An important methodological choice of our study is the ``neutral'' approach used for the selection of the variables within the financial statements. A part from removing related and redundant measures (total and subtotals), we used all the available information applying the same weight to each variable in the vectors. This agnostic approach is in line with the goal of the paper, i.e. introducing the concept of Accounting Network, although we are aware that practitioners can give a different importance to each variable of the financial statement. In our perspective we expect that the relevant information will emerge in a bottom up process, as a spontaneous feature selection carried by our methodology.
Finally, we introduce a confidence level (95\%) during the link formation. By using a Montecarlo sampling test, if the cosine similarity is statistically significant with 95\% of confidence we retain the link otherwise we discard it. As a result of this filtering procedure, we observe that the networks tend to be very dense and almost complete. The most of the information is carried by the weights of the links and less by the simple topology (degrees and other structural features). 
 


\subsection*{Community detection}
A classical method to investigate the structure of a network is the search of communities, i.e. regions of the network with larger \textit{internal} links density. Intuitively, these regions are formed by clusters of nodes with higher degrees or, for weighted networks, with larger strengths. Several methods were proposed to find network communities without imposing a priori the number of communities but letting them emerging from the network itself. Among others we cite the optimization of the modularity that is a measure of how much the link structure differs from the random network where links are assigned with uniform probability and internal communities are not present (a part from fluctuation). For weighted networks, the modularity is defined by the following formula:\\
\begin{equation}
    Q_{w} = \frac{1}{2W} \cdot \sum_{ij} \left( w_{ij} - \frac{s_i s_j}{2W} \right) \delta (c_i,c_j)
\end{equation}
\noindent where $s_i = \sum_{j} w_{ij}$ and $s_j = \sum_{i} w_{ij}$ are the strengths (sum of weights) of the nodes $i$ and $j$ respectively, $W=\sum_{ij} w_{ij}$ is the total sum of the weights and the function $\delta (c_i, c_j)$ is equal to 1 if $(i,j)$ belong to the same community or 0 if they are members of different communities. The maximum modularity value is 1 (an ideal case for which the communities are isolated) and can also take negative values. The 0 value coincides with a single partition that will correspond to the whole graph. A negative value means that there is no particular advantage in separating the nodes in that particular clusters and so there is not community structure whatsoever.\\
\noindent To study the presence of communities it is often necessary to prune the network cutting the links if their weight is below a certain threshold. In our case we intend to consider only the links formed by nodes having a large similarity/weight $w$ of their financial statement vectors. The procedure of pruning can be guided by the use of the tools related to the community detection methodology (\cite{Fortunato:2010}). In particular, working with the modularity optimization function (\cite{Newman:2004}), with the Louvain technique (\cite{Blondel:2008}), it is possible to look at the \textit{significance} associated to the threshold (as in \cite{Traag2013}), where the modularity is introduced as a parameter to check for the best resolved community structure. We use this parameter to help finding a reasonable pruning threshold range of values for the networks. A rule of thumb in this process is indeed avoiding network fragmentation, i.e. keeping the graph connected while removing not significant links. We made extensive tests computing quality/significance of the partitions (looking at the modularity parameter) using different pruning thresholds (i.e. removing the links having a low weight), determining a range of weights thresholds ($0.35< w_{i,j}<0.5$) that helps to prune the original networks to an optimal level. In this interval, communities are stable and the interpretation of each region can be seen as a result of the financial statement similarities across banks in different countries. 

\subsection*{Network measures vs. economic indicators}
Comparisons among network measures and economic indicators are provided to describe the correlation between nodes' network topology and economic behavior. We study these features by means of extensive linear correlation tests (Pearson correlation) for the overall set of banks for each year and we verify the significance of the estimates by means of parametric tests. These estimates are based on the filtered networks, which are themselves based on the significance and the quality of the community detection algorithm. This analysis shows how nodes' network properties (e.g. \textit{Strength} or \textit{Clustering Coefficient}) are associated to basic economic indicators (e.g. \textit{Return on Assets}, \textit{Total Assets} and \textit{Total Debts to Total Assets}), thus showing whether nodes' topological properties are positively or negatively related to certain economic features and how these relationships have weakened or reinforced during the crisis.\\
\noindent Clustering coefficient is a measure of the local tendency of the nodes to form small regions of fully connected nodes, it is an average measure of the local clustering coefficient (actual number of triangles centered in each node over the total). Return on assets (ROA) is the net income over total assets and is a measure of the bank performance. Total debts to total assets is an indicator of the leverage of the bank and it is computed as the ratio between debts and its size (measured by total assets).

\subsection*{Principal Component Analysis}
Once communities are identified, we attempt to describe which financial statement variables are more likely to characterise these clusters. In order to facilitate comparability, we focus on those indicators more popular within the set of variables utilised to compute the cosine similarities (i.e. those indicators appearing with larger frequency in the entire dataset). In fact the inclusion of very poorly represented measures across different banks would have made the comparisons less effective with potential biases related to e.g. different regulations frameworks or geographical memberships. Hence, since we are interested in disentangling potential similarities/peculiarities across different communities, we prefer to rely on common and well-diffused measures of banking activities among those present in banks' financial statements. In addition, we enrich this set by means of indicators such as ratios (e.g. \textit{Return on cap} and \textit{Total debts to total assets}) and aggregated measures (e.g. \textit{Total assets}). Community detection identifies four main clusters, whose constituents are more numerous and stable in time. For the sake of conciseness, the \textit{Result} section will focus mainly on these communities. In particular, for each year we describe by means of Principal Component Analysis (PCA) which economic features are more (less) able to contribute to the explained variability of communities' members.\\ 
PCA is a multivariate technique that analyses observations described by several inter-correlated variables. PCA extracts the important information from the data and expresses it as a set of new orthogonal variables (principal components). In our exercise, since measures present different ranges of dispersion (e.g. by construction some ratios are bounded) we rely on a scaled version of PCA; finally, we consider only principal components with eigenvalues greater than $1$ (in almost all cases they correspond to the first $3$ components). Then, we compute the proportion of the variance of each original economic measure that can be explained by the selected principal components. This, in turn, leads to a ranking of the original economic measures in terms of their ability to describe a certain community's variability. In particular, since we are interested in how the onset of financial crisis has affected the banking system, we split this analysis in three periods: from $2001$ to $2006$ (before the crisis), from $2007$ to $2009$ (the onset of the crisis), and from $2010$ to $2013$ (after the breakdown of the markets). For each period we decided to characterise each community by the top and the bottom three measures, thus analysing how these ranks have evolved over time and across communities.

\section*{Results}
This section shows how Accounting Networks represent a complementary technique to traditional financial networks for the study of the banking system.\\
While financial networks reflect the view from the market, related to e.g. the pairwise correlations of stock prices, Accounting Networks capture the effects of business decisions on financial statements measures and on business models of different institutions. An ``ideal" investigation of the financial system would involve also a detailed analysis of the money flows among companies, which determine the so called ``mutual exposures" (an important contagion channel). Unfortunately, these high granular and detailed data are usually not available. However, financial statements provide an aggregated view of mutual exposures and obligations for different maturities and types of instrument. This is an important point in favour of Accounting Networks as they report summarised information for e.g. phenomena occurring with different time scales and contractual terms, as opposite to the financial networks that rely only on homogeneous (daily or intraday) market data.\\

\subsection*{Community Detection Results}

\begin{figure}[htbp]
\centerline{
\includegraphics[scale=0.25]{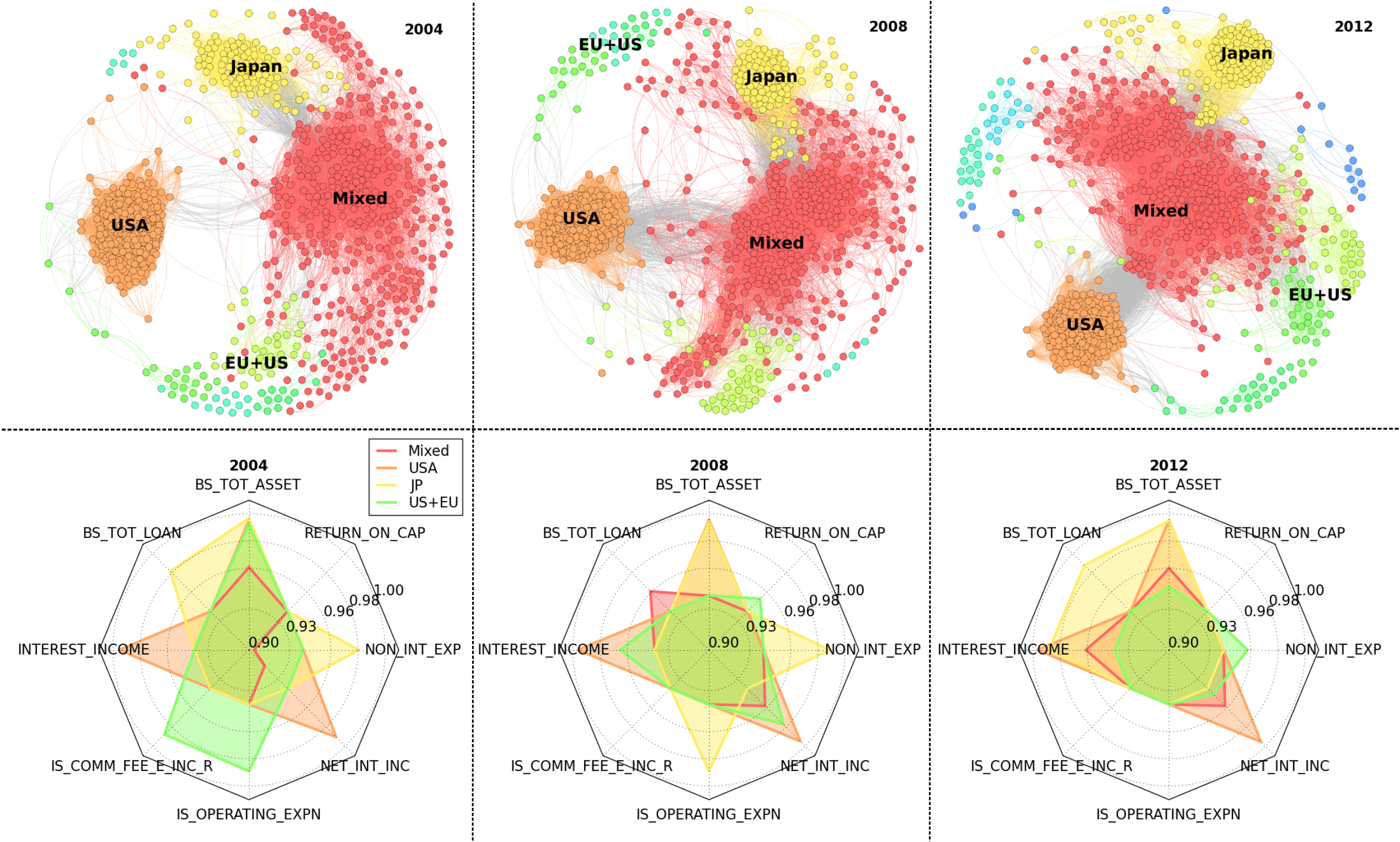}
}
\caption{In the upper panels it is shown the Community Structure for the three periods. The impact of the financial down-turn of 2007-08 seems to be reflected more heavily after the crisis, with the emergence of many sub-region communities as a response against the deteriorated market conditions. In the lower panel the most important financial statements components by the PCA analysis.}
\label{figCD_PCA}
\end{figure}

\noindent In this sub-section we focus our attention on the bottom up clusterization of the network from the application of the community detection algorithm and on the presence of geographical structures arising when we label each bank with its country. Therefore, we describe whether banks belonging to different countries (as a proxy for different regulations and/or level playing fields) have shown the tendency to be part of separate or, alternatively, common clusters and we verify, by analysing communities' evolution over time, whether the crisis influenced these configurations. In particular, our community detection analysis on Accounting Networks shows these main results.\\
It exhibits the presence over time of a clear community representing US banks and another one composed by Japanese banks, although for both regions there is also an additional smaller second group quite persistent in time. By contrast, it is not possible to identify a single and an unambiguous European community, since banks belonging to European countries seem to be likely to form national or sub-regional communities or to be included in a vast and geographically heterogeneous cluster (hereinafter the \textit{Mixed} community). In addition, Asian banks are fragmented in several sub-regions where, in particular, the Arab and the Indian-Pakistan groups emerge. Therefore, the detection of communities within Accounting Networks reveals the presence of two homogeneous clusters corresponding to US and Japanese banks surrounded by a more diversified cloud of banks belonging to different countries; remarkably, European banks are not able to clusterise together in a single community, while it persists over time a certain level of separation based also on national borders. Hence, an interesting contribution of the paper points to the presence of a large and geographically heterogeneous community, which can be related to the fact that the globally established regulatory framework might have indeed accelerated the tendency of banking activities of different countries to converge into more uniform banking practices. This is shown for instance in Figure \ref{figCD_PCA} where we also observe that the outbreak of financial markets contributed to make the Mixed community more cohesive; furthermore, although still representing separate communities, both US and JP clusters result topologically closer to the Mixed community after the breakdown of 2007-08, thus supporting the interpretation of a gradual convergence of different areas into more similar patterns. In addition, the application of the community detection on Accounting Networks allows to identify even small communities, such as those related to African or Scandinavian banks. This represents a quite promising aspect of the methodology, since it ensures the detection of local reliable communities although the approach taken so far is eminently agnostic.\\
It is not simple to explain the reasons behind the emergence and evolution of these communities, however it is possible to advance some intuitions based on the impact of globally recognized accounting standards (\cite{FASB}), the establishment of supranational supervisory and regulatory authorities, and on the role of the harmonization process of banking practices which have been implemented through e.g. the various Basel regulations (\cite{BASEL}). These contributions point to a common level playing field, which might have facilitated the emergence of a large and geographically heterogeneous community and its increasing topological proximity to both US and JP clusters. However, latter communities highlight the persistence of regional peculiarities. In Japan a deregulation process, known as the 'Japanese Big Bang', was formulated during the 1990s to transform the traditional bank-centered system into a market-centered financial system characterised by more transparent and liberalised financial markets (\cite{JP1}). In fact, peculiar features of Japanese banking sector were the over-reliance on intermediated bank lending, the absence of a sufficient corporate bond market and a marginal role for non-bank financial institutions, whose main consequences were an abundance of non-performing loans, excess in liquidity, scarce investments and low banks profitability (see e.g. \cite{BATTEN}). Although this program was intended to cover the period 1996-2001, the goals have not yet been achieved and policy makers' continuing reform efforts to remove past practices by market participants confirm the slowing convergence of the Japanese regulatory framework to a capital-market based financial system (\cite{JP2}). Thus, the presence of the JP community which gradually tends to the Mixed cluster is in line with evidences from the Japanese financial sector reforms aimed to change its reliance on indirect finance into a system of direct finance related to capital markets. Furthermore, it is remarkable the presence of a US community quite stable over time, which seems to be progressively attracted by the Mixed cluster. The US financial system presents peculiar features compared to other geographical areas. It is characterised by a relatively greater role of capital market-based intermediation, a higher importance of the 'shadow banking system', and differences in the accounting standards (\cite{ECB}). The impact of non-bank financial intermediation relates to the use of originate-to-distribute lending models, which determine the direct issuance of asset-backed securities and the transfers of loans to government-sponsored enterprises (GSEs, e.g. Fannie Mae and Freddie Mac). Financial innovation played a key role and the increasing use of securitisation explains the low percentage of loans to households on banks' balance sheets (\cite{ECB}). In addition, the US 'shadow banking system' is highly dependent on the presence of finance companies, money market funds, hedge funds and investment funds, which influenced the growth of total assets in the US financial sector during the last decades (\cite{shin2012global},\cite{shadow}). The presence of a distinct community is probably also due to differences in accounting standards which mainly involve the treatment of derivatives positions between the US Generally Accepted Accounting Principles (US GAAP) and the International Financial Reporting Standards (IFRSs). In particular, US GAAP allows to report the net value of derivative positions with the same counterparty under the presence of a single master agreement, thus impacting on the size representation of balance sheets items. However, in Figure \ref{figCD_PCA} we observe that the US community (similarly to the JP community) is gradually approaching the Mixed community, and the consequences of the breakdown of 2007-08 seem to have enhanced this behaviour. Among the possible several reasons, it is worthwhile to consider the impacts of the reform on the OTC derivatives market (embedded in the Dodd-Frank Act) and the Basel III new banking regulation, which may have facilitated similarities among US institutions and their peers in the Mixed cluster.

\subsection*{Relationships between Economic Indicators and Network Properties}
\begin{figure}[ht!]
\centering
\includegraphics[scale=0.25]{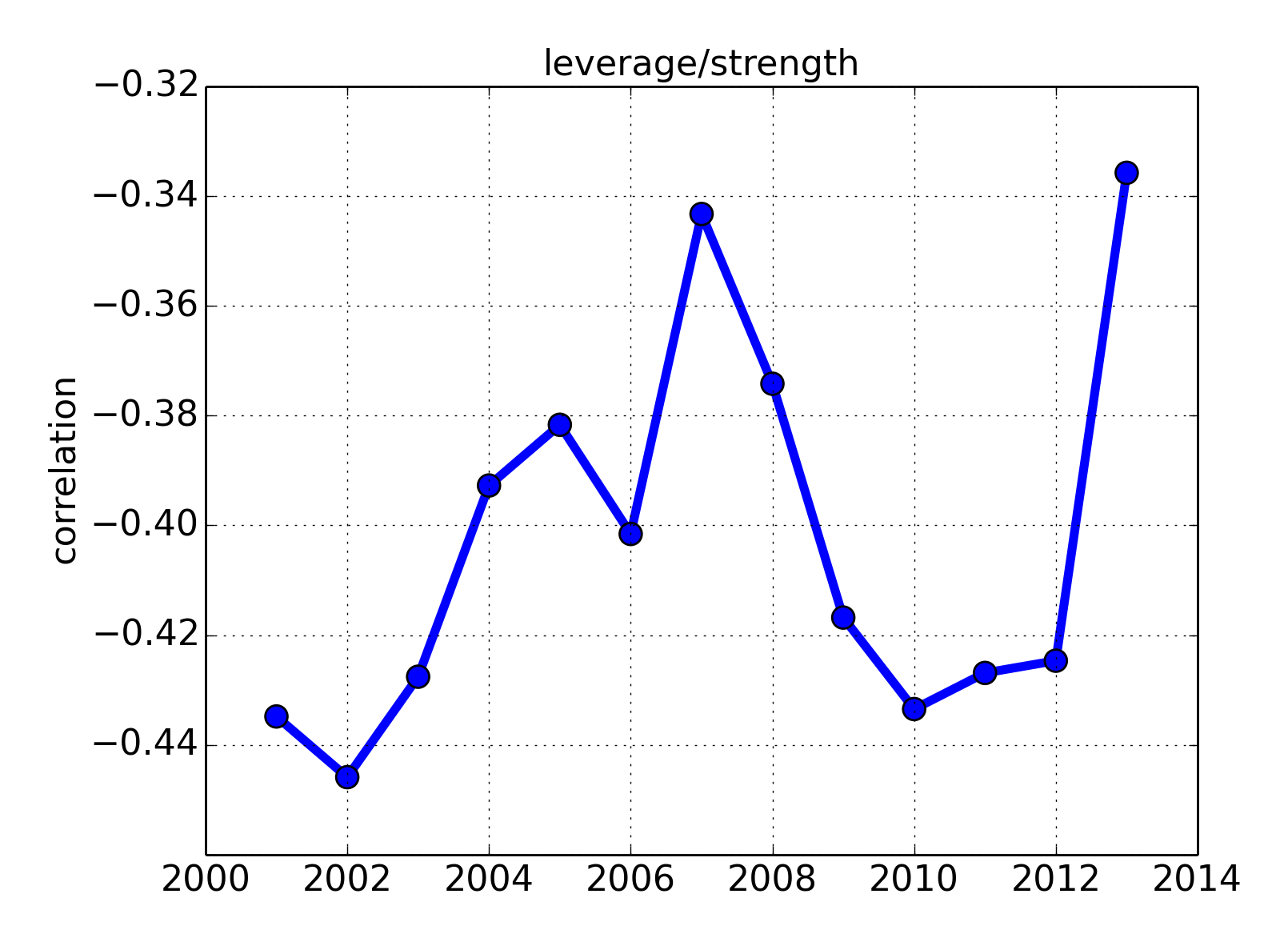}\hspace{-0.5cm}
\includegraphics[scale=0.25]{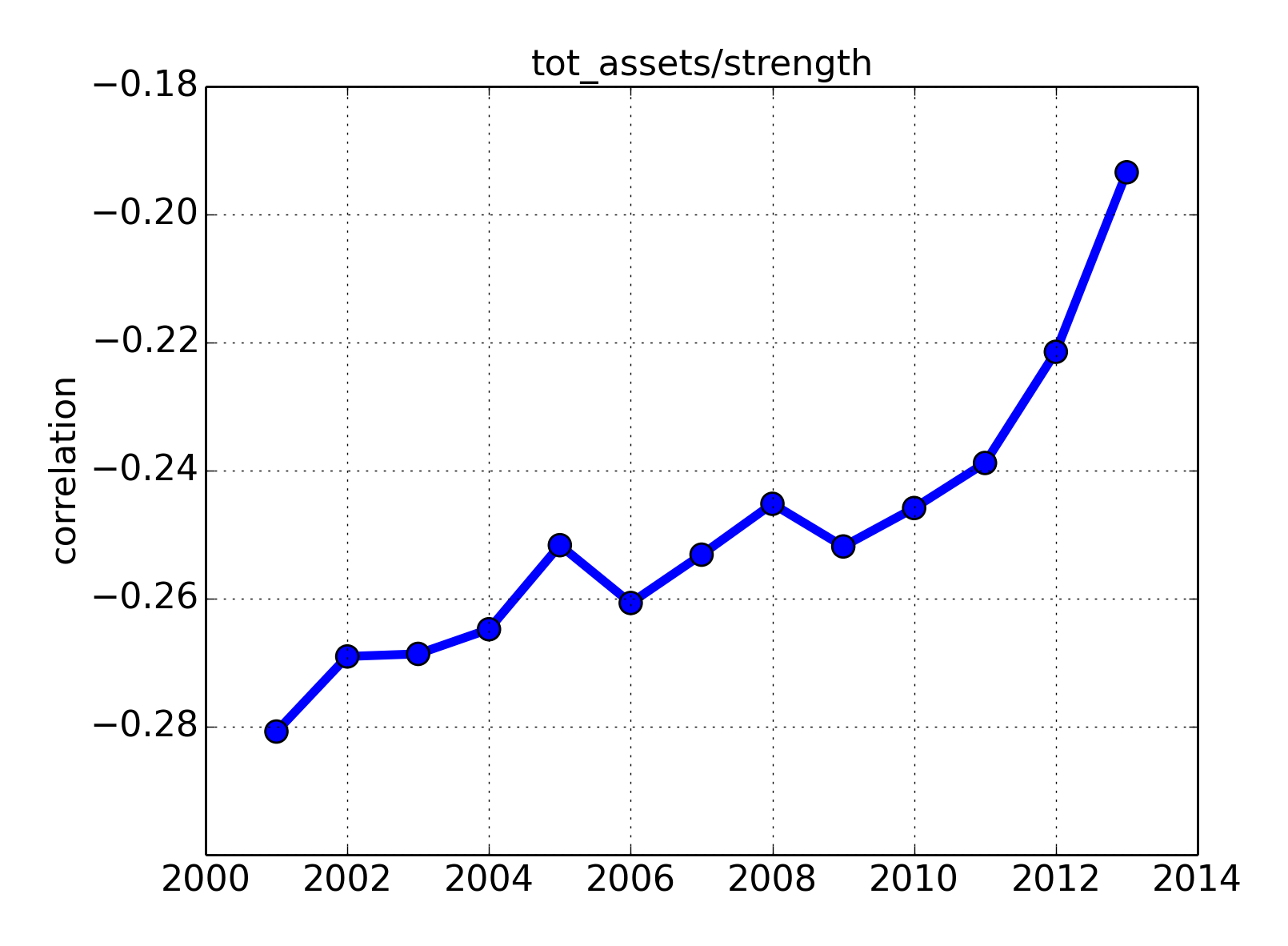}\hspace{-0.5cm}
\includegraphics[scale=0.25]{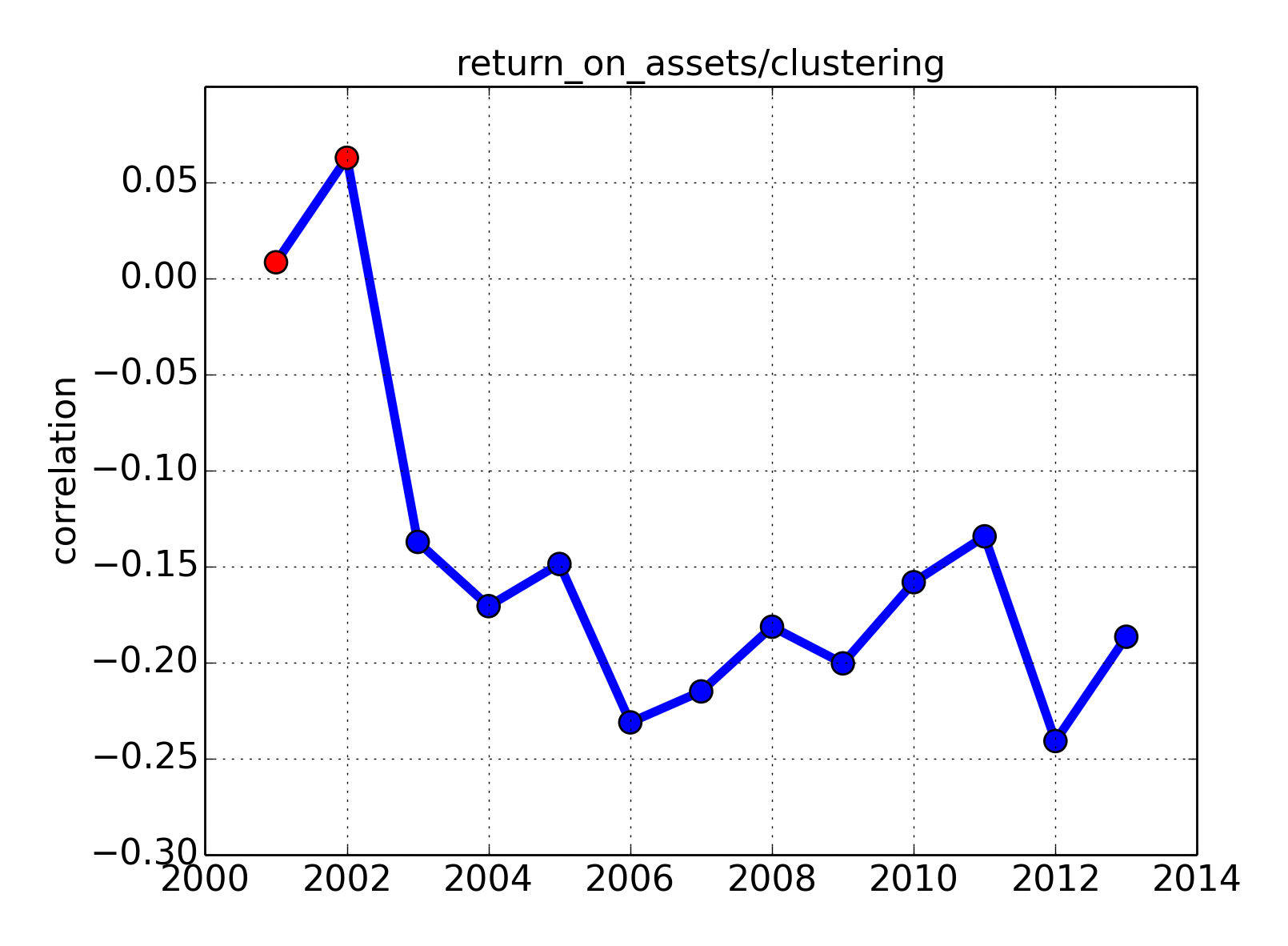}\hspace{-0.5cm}
\caption{In these plots we present the correlations between banks' Strength versus the Total Debts to Total Assets (Leverage) (plot on the left), Strength versus Total Assets (Size) (plot on the middle) and Clustering Coefficient versus Return on Assets (Performance) (plot on the right). The correlation is computed across the years 2001-13. It is clear the effect of the financial crisis across the outbreak of 2007-08. Red points stand for no-significant estimates, while blue points refer to significant estimates.}
\label{figCorrelations}
\end{figure}

\noindent In this Section we provide a preliminary investigation of the relationships between banks' economic indicators and their network properties. In order to characterise banks, we consider three common proxies for their classification: \textit{Return on Assets} (for the \textit{Performance}), \textit{Total Assets} (for the \textit{Size}) and \textit{Total Debts to Total Assets} (for the \textit{Leverage}). Then, comparisons are presented against two basic network measures: the \textit{Strength} and the \textit{Clustering Coefficient}. For each year from 2001 to 2013, we provide some insights for these relationships by estimating for the overall sample the correlations between banks' economic indicators and network measures. As explained in the Method, in this exercise we consider the network filtered according to the quality/significance of the \textit{Louvain} community detection algorithm. This helps us in the assessment of the significance of our results. Below, we show some examples to discuss how these relationships have evolved over time.\\
\noindent In particular, we investigate whether once the effects of the crisis have spread throughout the financial sector, the capacity of traditional economic indicators (e.g. leverage, size, performance) to group banks could result undermined. For instance, the onset of the financial crisis clearly affects the relationships between \textit{Total Debts to Total Assets} and network properties. Although the correlation between \textit{Strength} and \textit{Total Debts to Total Assets} remains negative during the entire sample period, the breakdown of financial markets seems to further enhance this effect for subsequent years (Figure \ref{figCorrelations}, plot on the left). Thus, this relationship suggests that, after the onset of the crisis, the use of leverage became on average more anti-correlated to the \textit{Strength}. This implies that banks that are more dissimilar in terms of their financial statements (i.e. with lower values of \textit{Strength}) are those that turned out to be less capitalised (i.e. with higher values of \textit{Total Debts to Total Assets}). Furthermore, one might be interested in understanding the role played by the \textit{Size} which represents a typical indicator utilised to classify banks. The correlation between \textit{Strength} and \textit{Total Assets} is almost flat and negative even after the collapse of 2007-08, but it shows an increasing trend in the recent period (Figure \ref{figCorrelations}, plot on the middle). Hence, it seems that after the outbreak of the crisis the \textit{Size} became less correlated to the similarity among banks, as estimates pointing sharply to zero seem to suggest. We finally analyse the relationship between \textit{Performance} and network properties (Figure \ref{figCorrelations}, plot on the right). In particular, in order to mimic how the presence/absence of more connected groups of banks is related to economic results we consider the \textit{Clustering Coefficient} for determining the level of structure in the system. Although poorly statistically significant in the early 2000s, correlations with \textit{Return on Assets} exhibit a decreasing pattern before the onset of the crisis and then remain negative although slightly erratic. The negative relationship between \textit{Clustering Coefficient} and \textit{Return on Assets} seems to suggest that the presence of well connected areas in the network (nodes with higher clustering coefficients) do not foster economic performance.\\
\noindent These basic examples suggest that a clear investigation on the relationships between economic indicators and network properties might be not always conclusive. Moreover, once we consider the entire set of banks, there might be some cases where estimates are poorly significant. Still, some remarkable effects arise from this investigation strategy and preliminary results point to a turning point in the correlations across the outbreak of the financial crisis. In particular, diagrams confirm that leverage is an useful indicator for differentiating banks, hence deviations to a lower capitalization are associated to increasing dissimilarity with the rest of the system and the impact of the crisis suggests a reinforcement in this relationship. By contrast, it seems that size does not contribute too much on the similarity between banks after the breakdown of 2007-08, while it played a greater role before and during the crisis. Finally, the relationship between performance and the structure of the system is less clear and prevents straightforward conclusions.\\
The identification of economic features potentially able to characterise specific portions of the system is addressed in the next sub-section.

\subsection*{PCA results}
Community detection shows the presence of three large clusters (Mixed, US, and JP) and an additional quite stable and persistent but smaller community (mostly US+EU banks). In this Section we provide a way to describe how these communities can be represented in terms of economic features (see Figure \ref{figCD_PCA}). Given the multi-dimensionality of the set of measures utilised to build the networks, we adopt a Principal Component Analysis approach to identify those measures which contribute more (less) to the explained variance within each community. For the sake of simplicity, we propose the ranking of the top (bottom) three measures for each community during the following intervals: pre-crisis ($2001-2006$), crisis ($2007-2009$), and post-crisis ($2010-2013$). In particular, for each year we compute the contribution of the original measures to explained variance; then, we average within each sub-period and we determine the rankings based on the mean period values. Below, we name the community with a mixed geographical composition as \textit{C0}, while we refer to the communities with a prevalence of US, JP and European plus US banks as \textit{C1}, \textit{C2} and \textit{C3}, respectively.\\
This representation allows us to compare communities' features over time and across different groups. For instance, we observe that \textit{Total Assets} and \textit{Interest Income} are quite frequent among top measures contributors, while \textit{Total Debts to Total Assets} is recurrent among measures in the bottom rankings. This is not surprising given banks heterogeneity in terms of the size (\textit{Total Assets}) and the economic results (\textit{Interest Income}) distributions, in contrast with the tight constraints on leverage (\textit{Total Debts to Total Assets}) due to regulatory requirements. By focusing on the top rankings we notice that \textit{C0} and \textit{C1} have fairly stable top contributors, while communities \textit{C2} and \textit{C3} are more affected by the wave of financial turmoil. Furthermore, bottom rankings seem to be on average only slightly influenced by the choice of different sub-periods. In addition, differences between mean values among the set of top three and the set of bottom three contributors are quite stable over time with only few exceptions, while the middle part of the distribution of measures' contributions (not reported, available from authors upon request) is in general quite sparse. For these reasons, we prefer to focus on the top and the bottom rankings to describe communities' features.\\
One might be interested in how the outbreak of financial crisis have influenced these rankings. Top composition of \textit{C1} is unaffected by the 2007-08 financial breakdown, while \textit{C0} is only partially modified by the onset of the crisis (\textit{Interest Income} is replaced by \textit{Net Interest Income}). Conversely, \textit{C2} presents a quite different configuration during the crisis sub-period when it exhibits a relevant role for expenses measures (i.e. Non Interest Expenses and Operating Expenses). Similarly, income statement measures become more relevant among top contributors also within the \textit{C3} community. Interestingly, community \textit{C0}, which is characterised by a mixed geographical composition, and the US community (\textit{C1}) reach identical top contributors after the outbreak of 2007-08, while the JP community (\textit{C2}), which shows the same top contributors as community \textit{C0} in the first sub-period, seems to react differently during the crisis, although in the third sub-period it shows again top contributors similar to \textit{C0} (and to \textit{C1}). By contrast, community \textit{C3} seems to present a peculiar pattern over time.\\
Therefore, the crisis sub-period coincides with remarkable differences in the top contributors, while the recent sub-period points to a renewed tendency to get similar contributors for a wider set of banks (\textit{C0} and \textit{C1}, and partially \textit{C2}). This seems to be in line with the above discussion on community detection results, where we highlighted a gradual proximity between clusters over time. Hence, these results suggest that heterogeneity within clusters is driven by similar economic measures after the crisis, although specific differences persist. This is the case for instance of loans, which are not present among top contributors in the US community while they are in the top ranking of both the Mixed and the JP community (as expected according to the above discussion). We also notice that the crisis seems to suggest an increasing importance of income statement measures in terms of contribution to the explained variance within communities. 
The breakdown of financial markets affected banks' results and this justifies the high level of heterogeneity expressed by income statements indicators. This can also be related to the impact of the crisis on financial statement measures and on the different ways banks update their balance sheet structures compared to the recognition of economic results as reported in the income statements items. Similar comparisons can involve also the bottom three measures, but for conciseness we omit this part. 

\section*{Discussion}
\noindent In this paper, we depict the banking system through banks' financial statements. Our main contribution is represented by the introduction of a methodology to exploit balance sheets and income statements data to construct Accounting Networks. We show some relationships between economic indicators and network properties, which might provide some new useful insights for banking classification practices. Having depicted some effects of the recent financial crisis by using a simple framework is an encouraging sign for further extensions. We rely on ``neutral'' and ``naive'' techniques to build the Accounting Networks. In particular, among common approaches usually applied to describe similarities concepts, we adopt one of the basic method, i.e. the cosine similarity. Future works can exploit more advanced methodologies. Moreover, our selection of variables utilised to compute cosine similarities assumes that each component has the same importance. This is quite a naive hypothesis, which could be enriched by measures discrimination based on economic literature and/or practitioners practices. Finally, for accounting reasons we limit our study on annual financial statements, while a more detailed description of the system might easily involve the use of quarterly data. Despite these simplifying assumptions, our approach has the merit of introducing a novelty in the debate on banking networks, and we believe that future improvements in the directions outlined above will enforce Accounting Networks' ability to describe the evolution of banking systems.

 
\section*{Acknowledgments}
In this paper we thank the financial support of the Italian project CRISISLAB and the support of the Linkalab Laboratory for its open discussions and precious suggestions.

\bibliography{biblio}

\section*{Figure Captions}


\newpage
\section*{Tables}
\begin{landscape}
\begin{table}
	\centering
{\scriptsize 	\begin{tabular}{llrlrlr}
		\hline
		Community & Top Measures 2001-06 & Values 2001-06 & Top Measures 2007-09 & Values 2007-09 & Top Measures 2010-13 & Values 2010-13 \\ 
		\hline
		C0 & BS\_TOT\_ASSET & 0.9695 & BS\_TOT\_LOAN & 0.9588 & INTEREST\_INCOME & 0.9635 \\ 
		C0 & BS\_TOT\_LOAN & 0.9257 & NET\_INT\_INC & 0.9529 & NET\_INT\_INC & 0.9620 \\ 
		C0 & INTEREST\_INCOME & 0.9228 & BS\_TOT\_ASSET & 0.9492 & BS\_TOT\_ASSET & 0.9537 \\ 
		C1 & INTEREST\_INCOME & 0.9955 & BS\_TOT\_ASSET & 0.9964 & INTEREST\_INCOME & 0.9968 \\ 
		C1 & BS\_TOT\_ASSET & 0.9951 & INTEREST\_INCOME & 0.9957 & NET\_INT\_INC & 0.9954 \\ 
		C1 & NET\_INT\_INC & 0.9917 & NET\_INT\_INC & 0.9933 & BS\_TOT\_ASSET & 0.9953 \\ 
		C2 & BS\_TOT\_ASSET & 0.9935 & BS\_TOT\_ASSET & 0.9927 & BS\_TOT\_ASSET & 0.9943 \\ 
		C2 & BS\_TOT\_LOAN & 0.9854 & NON\_INT\_EXP & 0.9886 & NON\_INT\_EXP & 0.9877 \\ 
		C2 & INTEREST\_INCOME & 0.9770 & IS\_OPERATING\_EXPN & 0.9883 & INTEREST\_INCOME & 0.9876 \\ 
		C3 & BS\_TOT\_ASSET & 0.9817 & INTEREST\_INCOME & 0.9678 & NON\_INT\_EXP & 0.9671 \\ 
		C3 & IS\_COMM\_AND\_FEE\_EARN\_INC\_REO & 0.9803 & NON\_INT\_EXP & 0.9670 & NET\_INT\_INC & 0.9621 \\ 
		C3 & NON\_INT\_EXP & 0.9800 & IS\_OPERATING\_EXPN & 0.9624 & IS\_OPERATING\_EXPN & 0.9564 \\ 
		\hline
	\end{tabular}
}
\vspace{2cm}

	\centering
{\scriptsize 		\begin{tabular}{llrlrlr}
			\hline
			Community & Bottom Measures 2001-06 & Values 2001-06 & Bottom Measures 2007-09 & Values 2007-09 & Bottom Measures 2010-13 & Values 2010-13 \\ 
			\hline
			C0 & BS\_LT\_BORROW & 0.7196 & BS\_LT\_BORROW & 0.6723 & BS\_LT\_BORROW & 0.7185 \\ 
			C0 & BS\_SH\_CAP\_AND\_APIC & 0.7131 & BS\_ST\_BORROW & 0.6511 & TOT\_DEBT\_TO\_TOT\_ASSET & 0.6659 \\ 
			C0 & TOT\_DEBT\_TO\_TOT\_ASSET & 0.4386 & TOT\_DEBT\_TO\_TOT\_ASSET & 0.5360 & BS\_ST\_BORROW & 0.6537 \\ 
			C1 & RETURN\_ON\_ASSET & 0.7006 & BS\_LT\_INVEST & 0.5799 & BS\_SH\_CAP\_AND\_APIC & 0.8151 \\ 
			C1 & INTERBANKING\_ASSETS & 0.4987 & INTERBANKING\_ASSETS & 0.5724 & BS\_LT\_INVEST & 0.7075 \\ 
			C1 & TOT\_DEBT\_TO\_TOT\_ASSET & 0.4941 & TOT\_DEBT\_TO\_TOT\_ASSET & 0.1366 & TOT\_DEBT\_TO\_TOT\_ASSET & 0.1762 \\ 
			C2 & INTERBANKING\_ASSETS & 0.8878 & RETURN\_ON\_CAP & 0.7911 & BS\_ST\_BORROW & 0.8892 \\ 
			C2 & TOT\_DEBT\_TO\_TOT\_ASSET & 0.6980 & BS\_SH\_CAP\_AND\_APIC & 0.7245 & TOT\_DEBT\_TO\_TOT\_ASSET & 0.7574 \\ 
			C2 & BS\_SH\_CAP\_AND\_APIC & 0.5491 & TOT\_DEBT\_TO\_TOT\_ASSET & 0.5702 & RETURN\_ON\_CAP & 0.7498 \\ 
			C3 & BS\_SH\_CAP\_AND\_APIC & 0.8124 & BS\_CASH\_NEAR\_CASH\_ITEM & 0.8004 & BS\_CASH\_NEAR\_CASH\_ITEM & 0.7045 \\ 
			C3 & RETURN\_ON\_ASSET & 0.8007 & BS\_NON\_PERFORM\_ASSET & 0.7707 & IS\_INT\_EXPENSES & 0.6626 \\ 
			C3 & TOT\_DEBT\_TO\_TOT\_ASSET & 0.6345 & TOT\_DEBT\_TO\_TOT\_ASSET & 0.5311 & TOT\_DEBT\_TO\_TOT\_ASSET & 0.6519 \\ 
			\hline
		\end{tabular}
		\caption{First table shows the sets of top three contributors for each community, while the second table shows the bottom three contributors. Values represent the contributions of original measures to the explained variances. Rankings refer to averaged values along each sub-period: 2001-06, 2007-09 and 2010-13.}
}	\end{table}

\end{landscape}

\end{document}